# An Idea to Increase the Security of EAP-MD5 Protocol Against Dictionary Attack

Behrooz Khadem[1*], Siavosh Abedi [2]   Isa Sa'adatyar [2]

[1]Assistant Professor,
[2] MSc. Secure Communication and Cryptography
Faculty of Information and Communication Engineering
IHCU, Tehran, Iran

Bkhadem@ihu.ac.ir     sia.abedi72@gmail.com     isasaadatyar@yahoo.com

**Abstract**

IEEE 802.1X is an international standard for Port-based Network Access Control which provides authentication for devices applicant of either local network or wireless local network. This standard defines the packing of EAP protocol on IEEE 802. In this standard, authentication protocols become a complementary part of network security. There is a variety in EAP family protocols, regarding their speed and security. One of the fastest of these protocols is EAP-MD5 which is the main subject of this paper. Moreover, in order to improve EAP-MD5 security, a series of attacks against it have been investigated. In this paper at first EAP-MD5 protocol is introduced briefly and a series of the dictionary attacks against it are described. Then, based on observed weaknesses, by proposing an appropriate idea while maintaining the speed of execution, its security against dictionary attack is improved.

## 1. Introduction

The authentication protocol is implementation of a process on an insecure communication channel to ensure the legit identity of different parties of a connection. One of the most common approaches for this is password-based authentication. In 1981, with Lamport providing a password-based authentication protocol, this protocol was introduced for the first time [1]. Since then many protocols have been presented [2-6], but each one of them has in turn been analyzed and evaluated be different researchers and different weaknesses have been observed [7-11]. Therefore research on achieving a secure authentication protocol has been conducted until now.

Family of EAP protocols is considered as an important category of authentication protocols. In fact, EAP is a general framework for authentication protocols and is widely used in wireless networks and point-to-point communications. This family consists of more than 40 protocols, each of which uses different speeds and security for different purposes [12]. One of the protocols of this family is EAP-MD5, which is taken into consideration due to its speed [13]. The protocol, of course, has no security authentication due to its lack of mutual authentication and weakness against dictionary attacks, and this is why this article, by presenting a simple proposal, attempts to correct the core of EAP-MD5 to make the implementation of the dictionary attack more difficult. Since the proposed improvement in this article is conducted on the core of the protocol (and not the protocol layers), therefore the network layers of the protocol are neglected in our investigations.

In this paper, at first, section 2 briefly describes the core of the EAP-MD5 protocol. In section 3, some dictionary attacks against this protocol are described. Then, in section 4, an idea is proposed to improve EAP-MD5 security against the dictionary attack. The summary of the results is presented in section 5.

## 2. Investigation of EAP-MD5 protocol

EAP-MD5 is a password-based and one-way authentication protocol, often used in wireless X802.1 networks and PPP connections [3]. This protocol consists of three components, the applicant, the identity authentication (WAP) and the server (RADIUS), in which only the identity of the applicant is authenticated (Figure 1). It is recommended that the password length in this protocol to be at least 16 bytes [14]. If the layers of the network and the formatting of the protocol are neglected, this protocol will become a simple exchange of information to authenticate the identity of the applicant. The sequential implementation of the protocol consists of 10 steps as follows.

1. The applicant sends the start request
2. Identity authenticator sends one byte ID to the applicant.

---

* Corresponding Author ,   Bkhadem@ihu.ac.ir

3. The applicant sends his username, which is his ID, to the identity authenticator.
4. The authenticator sends exactly the same information to the service provider (only its formatting structure is different).
5. At this point, the server will check whether this username is in the database or not. If available, it randomly generates a 128-bit challenge and sends along with ID+1 to the authenticator.
6. The authenticator sends exactly the same information to the applicant (only its formatting structure is different).
7. At this stage, the applicant submits the three ID+1 values and his password to their challenge together with a hash (MD5) of them and considers its name to the challenge response. He then sends this hash to the authenticator.
8. The authenticator sends exactly the same information to the server (only its formatting is different).
9. At this point, the server performs step 7 (he also has the challenge, password and ID). If the amount of the received hash is equal to the received response to challenge from applicant, the "accept" sign will be sent to authenticator, otherwise, "reject' sign is sent to the authenticator.
10. Authenticator sends exactly the same information to the applicant (only its formatting is different).

## 3. A review on dictionary attacks on the EAP-MD5

In short, in an attack, the adversary first eavesdrop the authentication process and uses the obtained values to execute the (offline) attack to get the password value. Since in the protocol the value of the identifier and the challenge are explicitly sent to the user, in addition to hash output, the adversary has two of three inputs of hash functiont and only doesn't have access to the password value. Therefore, the adversary can conduct a brute force attack using a limited dictionary (usually a pre-defined list with common passwords) to put each guess one by one for the password and obtain the hash, and if the hash value equals to the amount that the applicant sent in response to the challenge for authenticator, then the password is found; otherwise, the value of the next password is extracted from the dictionary and re-evaluated. The adversary repeats this process until applicant's password can be found. Keep in mind that the easier the password is, the easier this process will be.

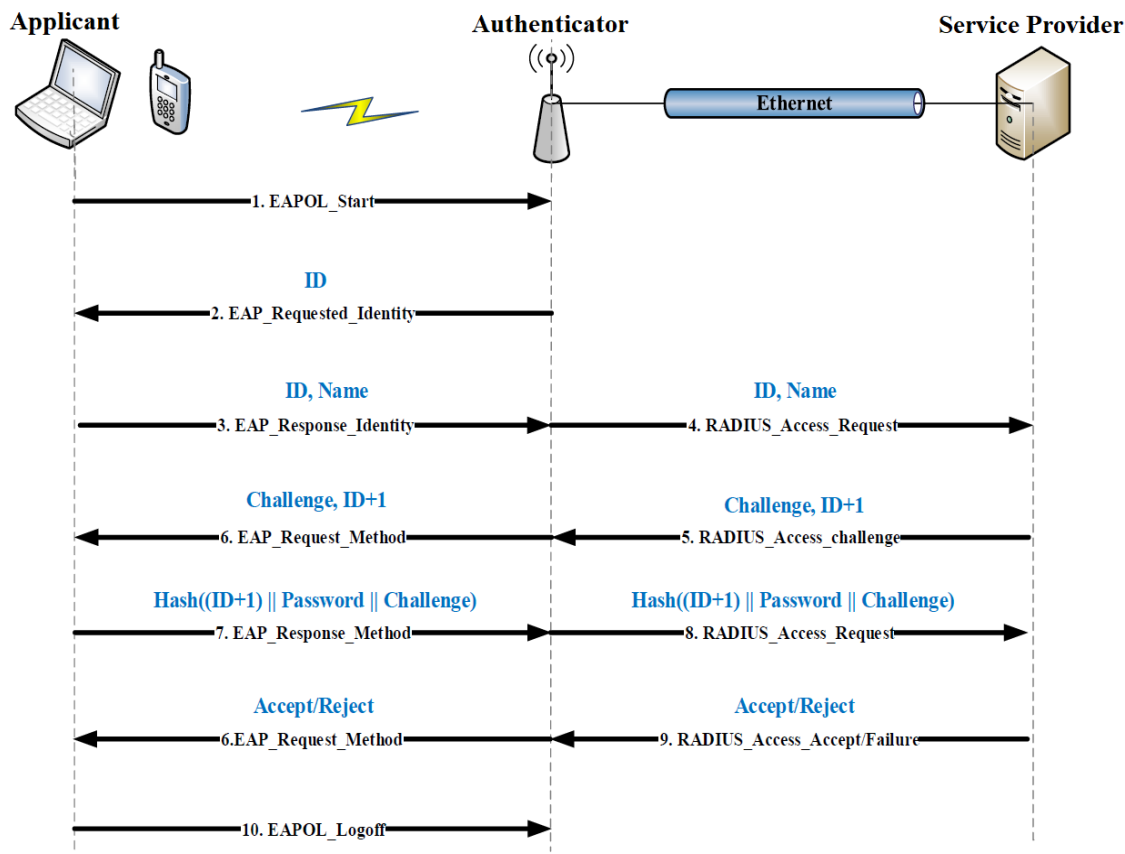

Figure 1 - Overview of the steps of the EAP-MD5 protocol

In 2008, Huang et al. in a wireless network, conducted a man-in-the-middle attack against the 802.1X standard and the EAP-MD5 protocol, and found the applicant's password [8].Again in 2008, Wright et al. managed to show that an application could find the password used in EAP-MD5 on a wireless network [15]. This application included an eavesdropping phase and a dictionary attack phase. The success rate of the attack depends on the type of dictionary used and simplicity of the password. Also, in 2012, Liu et al carried out an effective attack to recover the EAP-MD5 passwords in an IEEE X802.1 network and succeeded to find it in less than 3 days [9]. They found the password length at first using an online attack, and then using the rainbow table in the Hellman memory and time swap, were able to find the password value [16]. The proposed method in this article improves the security of the EAP-MD5 protocol against dictionary attacks [8, 9, 15], and makes them practically ineffective

## 4. Proposed method to improve EAPMD5

As seen in Section 3, one of the most successful attacks that can be made on EAP-MD5 is the dictionary attack for password guessing. In fact, the adversary uses a challenge response criterion for examining his own guesses about the password. The adversary guesses a password and, with some processing, can check the correctness of his guess. The reason for the simplicity of this process is that the adversary has access to two of three input parameters of the hash function (the applicant ID and the challenge of the server) because of their explicit transmission on the unsecure channel and so she can investigate its access by guessing the third parameter. So if the explicitly sent of one of these two parameters (due to the large length of the challenge, this parameter is more important) on the channel can be prevented, then the adversary will face more difficulty to guess the password. Here's an idea in which server challenge is not explicitly sent on the channel.

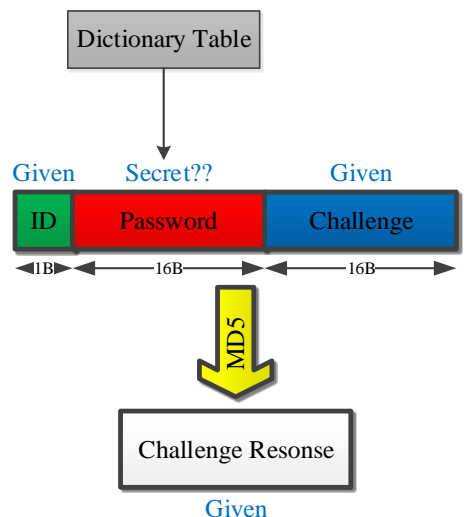

Figure 2 - An overview of the dictionary attack to find the password

In order to improve, the protocol steps changes as follows
1. The applicant will send the start request
2. Authenticator sends a one byte identity (ID) to the applicant.
3. The applicant sends his username (that is his ID) to the authenticator.
4. The authenticator sends exactly the same information to the server (only its formatting structure is different).
5. At this point, the server will check whether this username is in the database or not. If available, extracts his password and then generates a random 128-bit challenge and computes the request value according to (1) and sends it to the authenticator:

   Request = hash(ID $\oplus$ Challenge) $\oplus$ Password          (1)
6. The authenticator sends exactly the same information to the applicant (only its formatting structure is different).
7. At this stage, the applicant produces a time stamp and computes the response according to (2) and (3), and sends it along with the time stamp to the authenticator:
   C = Request $\oplus$ Password = hash(ID $\oplus$ Challenge)          (2)
   Response = hash(C $\oplus$ Time Stamp)          (3)
8. The authenticator sends exactly the same information to the server (its formatting is different only).

9. At this stage, the service first checks the timestamp and, if it is old, sends it "reject"; if the timestamp was updated, then the server will compute C according to (4). (It also has the challenge and ID); then uses it to compute response' according to (5); if the response' value is equal to the received value (response), 'accept' is sent to authenticator, and otherwise 'reject" is declared (Figure 3).

$$C' = hash(ID \oplus Challenge) \quad (4)$$
$$Response' = hash(C' \oplus Time\ Stamp) \quad (5)$$

10. The authenticator sends exactly the same information to the applicant (only the formatting structure is different).

As seen in figure 4, in this improvement, the service provider's challenge is not explicitly sent on the channel, which has increased the complexity of the dictionary attack.

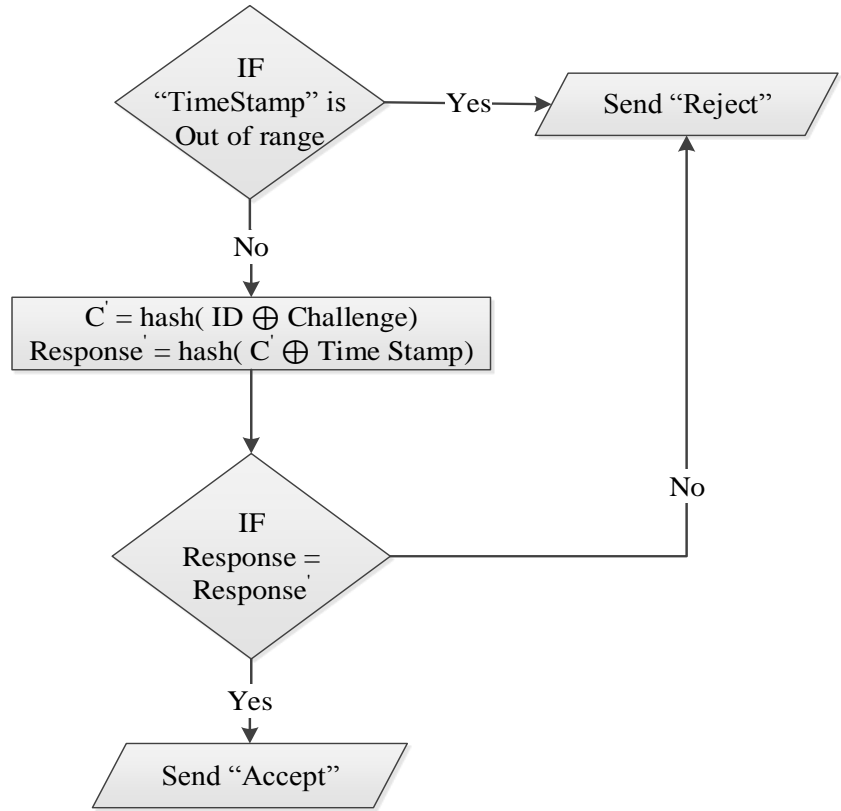

Figure 3- Flowchart of decision making procedure in the proposed protocol

As shown in Figure 4, in the proposed protocol, the challenge is not sent explicitly, and thus the adversary does not have two of the three input parameters of the hash function, and the actual guess of the password with the declared attack is ineffective.

## 5. Conclusion

The EAP-MD5 protocol, despite being fast and light, unfortunately, also faces some security problems. One of the security concerns of this protocol is its vulnerability to dictionary attack and quick search of password space by adversary. The main reason for this vulnerability is that in steps 5 and 6 of the main protocol, the value of server challenge is sent explicitly, so adversary can guess the password effectively by listening to this challenge. In this paper, a method is proposed that, while maintaining the speed of the implementation of the protocol (due to the use of a similar hash function), prevents the sending of a challenge explicitly to the channel and therefore effectively prevents the adversary from executing eavesdropping operations, therefore, effectively increase the complexity of implementing the dictionary attack for the adversary. Since the challenge is a random number, so if an adversary wants to conduct an attack on the proposed enhanced protocol, then, in addition to performing a dictionary attack, for each guessed password he should search the whole space of challenge too.

Since the length of the challenge is 128 bits, its search space is equal to $2^{128}$. In fact, the proposed protocol has added $2^{128}$ more complicated computing step to the complexity of the attack. In other words, if the complexity of an attack against the original protocol with a particular password is $2^c$ (the value of c depends on the adversary's computing utility), then the complexity of an attack against the proposed protocol in the same conditions would be equal to $2^{128+c}$. So if the $2^c$ value is assumed to be equal to 1, then the computational complexity of the $2^{128 \times 1}$ is also a very large value, conducting the attack for adversary requires a great deal of time and processing power. Table 1 shows the results of the comparison of the two methods. One of the other benefits of the proposed proposal is the duplication hash of information, which increase the complexity of the information exchanged. In addition, the timetable is also used in the proposal, which by itself also prevents replay attack.

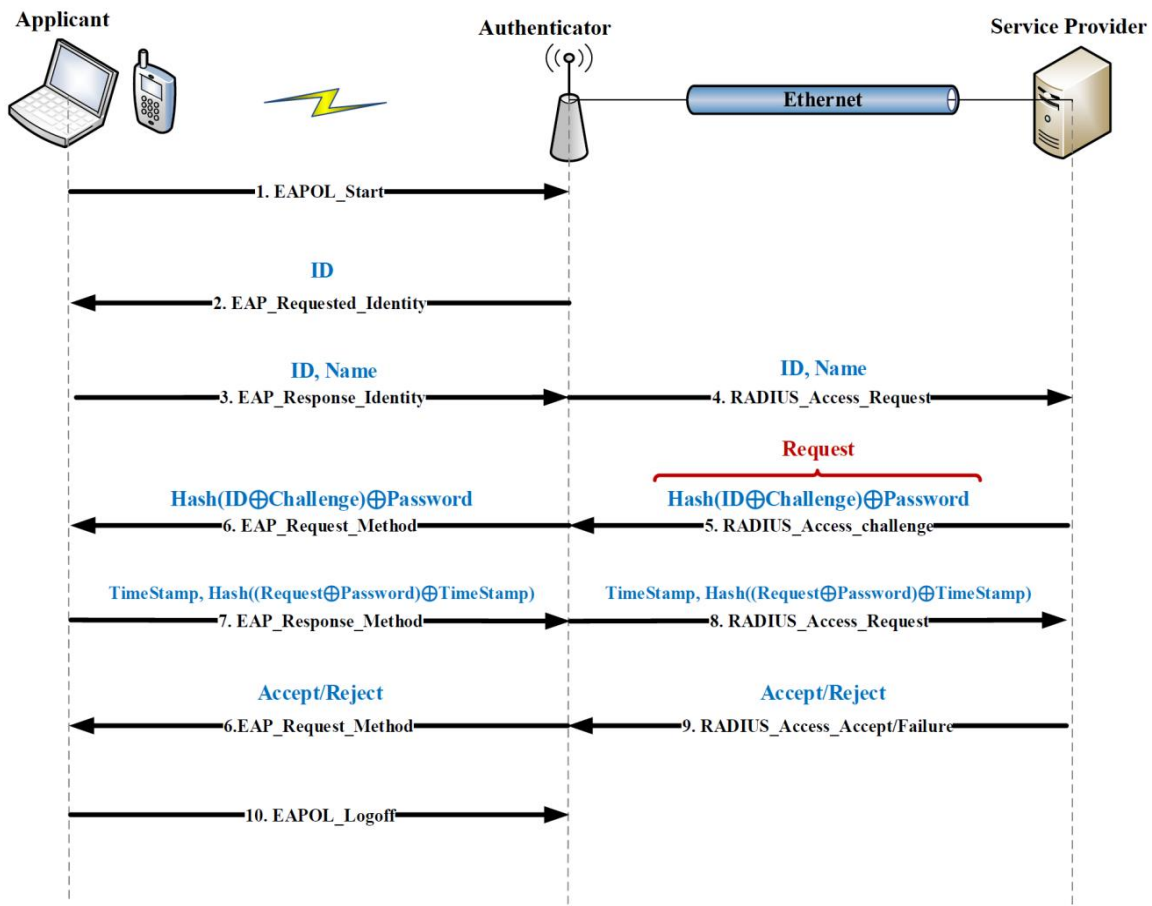

Figure 4 - Overview of proposed protocol steps to improve EAP-MD5

Table 1. Comparison of the main and proposed protocols

|  | Speed of the protocol | Robustness against reply attack | Computational complexity of attack |
|---|---|---|---|
| Main protocol | High | No | $2^c$ |
| Proposed protocol | High | Yes | $2^{128+c}$ |